\documentclass[journal=jacsat,manuscript=article]{achemso}

\usepackage[version=3]{mhchem} 
\usepackage{xcolor}
\usepackage{relsize}
\DeclareUnicodeCharacter{0308}{}

\author{Adam Yonge}
\author{Gabriel S. Gusm\~{a}o}
\affiliation[university1]
{College of Engineering, Georgia Institute of Technology, Atlanta, Georgia, USA}
\author{Rebecca Fushimi}
\affiliation[university2]
{Catalysis and Transient Kinetics Group, Idaho National Laboratory, Idaho Falls, Idaho, USA}
\author{A.J. Medford}
\affiliation[university1]
{College of Engineering, Georgia Institute of Technology, Atlanta, Georgia, USA}
\email{ajm@gatech.edu}

\title[An \textsf{achemso} demo]{Model-based design of temporal analysis of products (TAP) reactors: A simulated case study in oxidative propane dehydrogenation}

\abbreviations{}
\keywords{}

\newcommand*\rot{\rotatebox{90}}

\begin{document}

\begin{tocentry}

Some journals require a graphical entry for the Table of Contents.
This should be laid out ``print ready'' so that the sizing of the
text is correct.

Inside the \texttt{tocentry} environment, the font used is Helvetica
8\,pt, as required by \emph{Journal of the American Chemical
Society}.

The surrounding frame is 9\,cm by 3.5\,cm, which is the maximum
permitted for  \emph{Journal of the American Chemical Society}
graphical table of content entries. The box will not resize if the
content is too big: instead it will overflow the edge of the box.

This box and the associated title will always be printed on a
separate page at the end of the document.

\end{tocentry}

\begin{abstract}
 Temporal analysis of products (TAP) reactors enable experiments that probe numerous kinetic processes within a single set of experimental data through variations in pulse intensity, delay, or temperature. Selecting additional TAP experiments often involves arbitrary selection of reaction conditions or the use of chemical intuition. To make experiment selection in TAP more robust, we explore the efficacy of model-based design of experiments (MBDoE) for precision in TAP reactor kinetic modeling. We successfully applied this approach to a case study of synthetic oxidative propane dehydrogenation (OPDH) that involves pulses of propane and oxygen. We found that experiments identified as optimal through the MBDoE for precision generally reduce parameter uncertainties to a higher degree than alternative experiments. The performance of MBDoE for model divergence was also explored for OPDH, with the relevant active sites (catalyst structure) being unknown. An experiment that maximized the divergence between the three proposed mechanisms was identified and led to clear mechanism discrimination. However, re-optimization of kinetic parameters eliminated the ability to discriminate. The findings yield insight into the prospects and limitations of MBDoE for TAP and transient kinetic experiments.

\begin{figure}[H]
    \centering 
  \includegraphics[width=12cm]{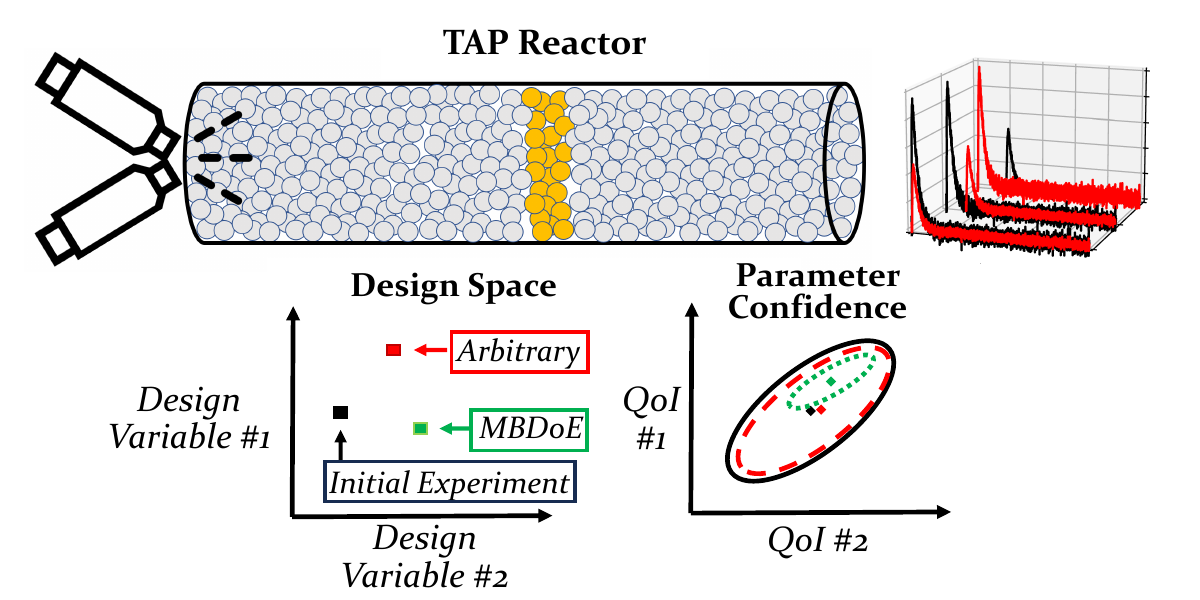}
  \caption{Visual Abstract.}
  \label{opdh_initial_fit_precision}
\end{figure}

\end{abstract}
\newpage

\section{Introduction}

Although heterogeneous catalysts have been studied for well over a century, there are many remaining challenges to systematically understanding the reaction mechanisms and active sites that govern their behavior in chemical reactors \cite{chorkendorff2017concepts}. These challenges are the result of many factors, including the complexity of materials, the various length scales involved in catalytic processes, and the time scales over which catalytic events take place \cite{wachs2005recent,salciccioli2011review}. By quantitatively understanding materials and their associated kinetics, we can rationally identify and optimize promising catalysts. A wide range of computational and experimental methods have been developed to explore the kinetics of heterogeneous catalysts. At the atomic scale, the electronic structure of individual atoms and molecules can be investigated with computational simulations, while surface science experiments are commonly used to study well-defined surfaces, and catalyst particles or packed-bed reactors (PBRs) can be utilized to provide insight into the reactor-scale behavior of catalysts. However, the ``pressure gap'' and ``materials gap'' create challenges in connecting behavior between these scales \cite{lu2021investigating,oyedeji2022cfd,norskov2011density,van2009molecular, reece2018crossing,reece2019dissecting,reece2021moving, medford2018extracting}. 

Even with the availability of these tools, it can be challenging to derive strong mechanistic insights into catalytic materials due to the presence of experimental, parametric, and structural uncertainties \cite{kennedy2001bayesian}. These uncertainties can easily lead to confidence intervals of quantities of interest (QoI) that span several orders of magnitude \cite{walker2016uncertainty}. For this reason, much effort has been made over the last decade to identify and reduce these sources of uncertainty. For example, Heyden and co-workers have propagated DFT free energy uncertainties, which are known to be  sensitive to the choice of exchange-correlation functional, to the QoIs in microkinetic models (e.g., turnover frequencies, apparent activation energy) to help discriminate between competing proposed reaction pathways\cite{walker2016uncertainty, walker2018identifying}.
N\o{}rskov and co-workers have also investigated the role of exchange-correlation uncertainty in catalyst screening by propagating the model uncertainties from the BEEF-vdW functional to microkinetic models for screening catalysts for the ammonia synthesis and syngas conversion reactions \cite{medford2014assessing, Schumann2018}. Vlachos and co-workers have similarly progressed the field through their study of scaling relationship errors on selectivity predictions, as well as uncertainty-based analysis of experimental data and coverage effects \cite{sutton2016effects, chen2021experimental,ulissi2011effect}. From these studies and others, various formalisms and software packages have also been developed to quantify and evaluate the impact of uncertainty within the field of catalysis and reaction modeling \cite{savara2020chekipeuq,walker2020chekipeuq,cohen2021chemical,wang2021nextorch}.
 
\begin{figure}[H]
    \centering 
  \includegraphics[width=10cm]{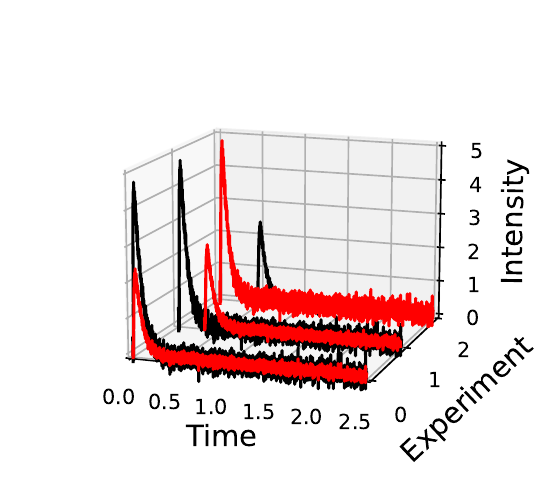}
  \caption{Illustration of pump-probe TAP experiments, where pulses of two reactants (red, black) are pulsed with varying intensities and delays between different experiments.}
\label{pump-probe}
\end{figure}

One challenge in catalysis is that it is often difficult or impossible to directly extract intrinsic kinetic parameters from experimental datasets. Steady-state kinetic measurements are typically sensitive to only a few parameters and are therefore known to be prone to overfitting when complex kinetic models are used \cite{Rangarajan_2017,dumesic1993microkinetics}. Surface science techniques allow the measurement of specific adsorption energies and reaction barriers, but they typically require well-defined surfaces that may differ from complex nanoparticles used in real applications, and the adsorption and reaction barriers that are extracted are also prone to uncertainty comparable to calculated quantities \cite{imbihl2007bridging,freund2001bridging,hammer2000theoretical}. Transient kinetic experiments can overcome some of these challenges because they are sensitive to a larger number of elementary processes compared to steady-state experimental approaches. In recent years, there has been renewed interest in the use of transient experiments to explore catalytic materials, including steady-state isotopic transient kinetic analysis (SSITKA) and  spectrokinetics \cite{vasiliades2020effect,lorito2022controlling,rezvani2020co2,vasiliades2019effect,moncada2018developing,bregante2021dioxygen,müller2017applications}. An additional transient kinetic experimental approach is the temporal analysis of products (TAP) reactor. The TAP approach uses a small (approximately 4 cm long) PBR operating under ultra-high vacuum (UHV) conditions. A TAP experiment consists of a series of rapid nanomolar pulses of reactants, closely related to molecular beam experiments \cite{libuda2005molecular}. A pulse valve introduces these molecular pulses at the entrance of the reactor, and the molecules diffuse to the outlet where a mass spectrometer is located to detect the outlet flux of all gasses. The reactor can be used with complex industrial catalysts and exhibits a well-defined transport regime, i.e. Knudsen diffusion, which aids in the deconvolution of transport and kinetics \cite{morgan2017forty,yablonsky2003temporal,gleaves1988temporal,redekop2022truth}. The TAP approach has also been used for pump-probe experiments, which can provide even more control over which elementary steps are probed by pulsing different reactants with time delays \cite{morgan2017forty,perez2007mechanism,kumar2010isotopic,wang2021understanding}. Altering the reactant feed varies the surface coverage, which, in turn, can alter the activity of various elementary processes. An illustration of the experimental conditions used in a pump-probe TAP experiment is shown in Figure \ref{pump-probe}. While TAP experiments offer information-rich data sets, complex data analysis and model fitting techniques are required to extract kinetic information, and the resulting models and parameters still have a broad range of uncertainties associated with them\cite{roelant2007noise,yonge2021tapsolver,yonge2022quantifying}. 

Gathering additional experimental data is one route to reduce uncertainty in fitted models, and this is well suited to TAP experiments, where pulses can be collected on the second time scale (with up to ten thousand pulses at different conditions collected within a single day), and experimental conditions can be relatively easily varied \cite{franceschini2008model}.
For example, Wang et al. used fixed variations in the pulse delay of isotopic oxygen species to better understand the active oxygen species in the oxidative coupling of methane process \cite{wang2022mechanistic}. These pump-probe experiments can help target the expression of elementary processes that might be masked under alternative conditions. At present, additional TAP experiments are typically selected on the basis of a user's chemical intuition, which is a qualitative approach that may not yield optimal results. 

The model-based design of experiments (MBDoE) has been studied for the last few decades and offers a quantitative means to guide experimental selection based on improved precision and mechanism discrimination \cite{franceschini2008model}. MBDoE can also help address issues with respect to structural uncertainty, which arises due to uncertainty in the elementary steps and the number of active sites involved in the reaction mechanism \cite{wachs2005determination,jain2013commentary,ong2013python}.  MBDoE has been applied to some transient kinetic examples, but the complexity of the reaction mechanisms and reactor models studied is less than that of typical TAP experiments \cite{waldron2019autonomous,quaglio2019online,waldron2020model}. 

In this study, we outline the theoretical terms of MBDoE in terms of TAP reactor experiments and apply the methodology to a synthetic oxidative propane dehydrogenation (OPDH) process. We identify potential limitations of this process and show how it can be modified to target the reduction of parametric uncertainty on kinetic parameters. We also explore the structural uncertainties present in kinetic mechanisms and active site structures through MBDoE for divergence. Some studies of mechanism discrimination have previously been performed for TAP experiments, but no implementations were introduced to quantitatively select between them \cite{kondratenko2006mechanism}. Mechanistic uncertainty is also explored through active site configuration variations in the synthetic OPDH process. The challenges of pairing mechanism discrimination with optimization are discussed, as well as prospects for making the MBDoE for precision and discrimination more practical.

\section{Methodology}

The equations associated with the modeling of the TAP reactor and MBDoE have been thoroughly outlined in previous publications \cite{yonge2021tapsolver, franceschini2008model}. For clarity, the necessary equations for this work are introduced in the following subsections. The proposed workflows for precision and discrimination are also introduced, as well as the OPDH mechanisms used as a case study. All simulations and optimizations were performed using the open source Python package TAPsolver \cite{yonge2021tapsolver}.

\subsection{TAP PDEs and Uncertainty Quantification}
The fundamental TAP reactor equation consists of a (Knudsen) diffusion and a reaction term (generalizing all reactions involving the gas species). This equation is written as:

\begin{equation}
\varepsilon\dot{\mathbf{c}}(x,t)-\mathbf{d}\circ\mathbf{c}''(x,t)=\mathbf{M}\times\mathbf{r}(\mathbf{c})\label{PDE-form-gas}
\end{equation}
where the void fraction is defined as $\varepsilon{}$, the concentrations of species $i$ in the system are defined by vector $\mathbf{c}=[c_i]^T$, with $\partial_t\mathbf{c}=\dot{\mathbf{c}}$ and $\partial_x\mathbf{c}=\mathbf{c}'$ representing the time (t) and spatial derivatives (x) of the concentrations, respectively, $\mathbf{d}=[D_i]^T$ defines the Knudsen diffusivity for each gas, $\mathbf{M}$ represents the stoichiometry matrix of the reaction system, $\mathbf{r}$ is a vector with rates of individual reactions, and $\circ$ represents the element-wise product. Surface species in the catalyst zone of the TAP reactor, represented by $u_i$, follow a similar equation, but without diffusive transport. As noted in the introduction, the outlet flux of each gas is the primary experimental data extracted from the TAP reactor. The outlet flux is defined as:

\begin{equation}
\mathbf{f}=\mathbf{d}\circ\mathbf{c}''(L,t) \label{flux_eq}
\end{equation}

where $\mathbf{f}$ is the vector of fluxes and $L$ is the length of the reactor. When fitting TAP experimental data with PDE solutions, it is also necessary to define the objective function over the $j$ time steps:
 
\begin{equation}
J^{N} = \frac{1}{2}\mathlarger{\sum^{\tau}_{j=1}} 
\boldsymbol{\varepsilon}_{j}^T\boldsymbol{\Omega}_j\boldsymbol{\varepsilon}_{j}
\label{objectiv_function_std}
\end{equation}
where $\boldsymbol{\Omega}_j$ is the precision matrix, i.e. inverse of the covariance matrix $\boldsymbol{\Sigma}_j$, and $\boldsymbol{\varepsilon}_{j}$ is the model residual:
\begin{equation}
\boldsymbol{\Omega}_j=\boldsymbol{\Sigma}_j^{-1}=\operatorname{diag}(\hat{\boldsymbol{\sigma}}^2_j)^{-1}\label{noise_diagonal}
\end{equation}
\begin{equation}
\boldsymbol{\varepsilon}_{j}=\hat{\mathbf{f}}_j-\mathbf{f}
\end{equation}
with $\hat{\boldsymbol{\sigma}}_j$ being the standard deviation of the experimental noise at time $j$.

The noise present in the observed data comes from the mass spectrometer and translates into uncertainty in the kinetic parameters. This standard deviation is defined in Equation \ref{objectiv_function_std} as $\mathbf{\hat{\sigma{}}_{j}}$, and is approximated as Gaussian, although the standard deviation depends on both time and gas species so that, in principle, it is possible to account for heteroscedastic errors that have been reported for data from mass spectrometers \cite{roelant2007noise}, though we assume homoscedastic errors in this work. Near the local minima of this objective, the shape of the well can be approximated as quadratic. Calculating the Hessian, which is defined as:

\begin{equation}
\mathbf{H}^{N}_{\boldsymbol{\theta}} =\nabla^2_{\boldsymbol{\theta}} J^N =  \frac{ \partial{}^{2} J^{N} } {\partial{\theta_k}\partial{\theta_l}} \label{hess1}
\end{equation}

near this point allows confidence intervals to be extracted following optimization \cite{oehlert1992note,cox2005delta}. In Equation \ref{hess1}, $\theta{}$ is the kinetic parameter of interest. The confidence intervals can then be extracted from this Hessian through the following equations:
\begin{equation}\label{hess2}
    \boldsymbol{\Sigma}^{N}_{\boldsymbol{\theta}} \approx [\mathbf{H}^{N}_{\boldsymbol{\theta}}]^{-1}
\end{equation}
\begin{equation}\label{hess3}
    \boldsymbol{\sigma} \approx \sqrt{\operatorname{diag}\left(\boldsymbol{\Sigma}^{N}_{\theta}\right)}
\end{equation}
where  $\boldsymbol{\Sigma}{}^{N}_{\theta}$ is the covariance matrix and  $\boldsymbol{\sigma}$ is the vector of standard errors of each fitted parameter due to experimental signal noise.

\subsection{Synthetic oxidative propane dehydrogenation case study}

We select OPDH as a case study since it is industrially relevant for propylene production, requires understanding of both rate and selectivity, and has mechanistic and active site complexity that provide a rich set of theoretical challenges \cite{gambo2021catalyst,cavani2007oxidative,carrero2014critical}.

The kinetics of the OPDH process have been broadly studied. We base the free energies of activation and reaction for our synthetic reaction mechanism on a combination of commonly observed values and the values found in the work by Chen et al\cite{chen2000kinetics,routray2004oxidative}. The mechanism used is provided in Table \ref{opdh_model_1}. The mechanism is not composed of elementary steps, which limits the complexity of the analysis and is consistent with the fact that not all elementary processes are likely to be observable from TAP experiments.
For example, the reverse reactions in a combustion reaction or some re-adsorption processes are unlikely to be observed, even with the low pulse intensity and pressure found in TAP. Similar reaction steps to those found in Table \ref{opdh_model_1} have been used in kinetic models for oxidation reactions of other hydrocarbons \cite{stegelmann2004microkinetic}.
 
We also explore the structural uncertainty in this case study, referring to both the uncertainty in the elementary steps involved and the active site(s) on which reactions take place. In the case of OPDH, it is often hypothesized that different reactions occur on different types of active sites \cite{fricke2022propane,chen2000kinetics}. For this reason, we defined two additional multi-site variations of the single-site OPDH mechanism (presented in Table \ref{opdh_model_1}). These reactions have two separate active sites, where mechanism 2 and 3 involve isolation of oxygen adsorption and propane combustion, respectively, as inspired by various hypotheses in the literature \cite{grasselli2005selectivity}. Notably, this presents a particularly difficult challenge in model discrimination, since the elementary steps included and the kinetic parameters are the same between all three models, allowing a specific focus on the question of whether TAP experiments can distinguish between single-site and multi-site mechanisms.

\begin{table}[H]
\begin{center}
\small\addtolength{\tabcolsep}{-5pt}
\begin{tabular}{c@{\hskip 0.15in}c@{\hskip 0.15in}c@{\hskip 0.2in}c@{\hskip 0.15in}c@{\hskip 0.15in}}
\caption{The Gibbs free energies of reaction and activation for the synthetic OPDH model. In mechanism 1, we include only a single active site for all kinetic processes. In mechanism 2, a separate site is included for oxygen (i.e. oxygen and the carbon species do not compete for the same site). In mechanism 3, surface hydrocarbons are allowed to combust on a separate active site from the remaining kinetic steps.}\label{opdh_model_1} \\
\hline  & Step & Reaction & $\Delta{}G$ (eV) & $G^{\ddagger}$ (eV) \\ 
\hline
& 1 & $C_{3}H_{8} + * \leftrightarrow{}  C_{3}H_{8}*$ & -0.2 & 0.3 \\
& 2 & $O_{2} + 2* \leftrightarrow{}  2O*$ & -0.7 & 1.25 \\
& 3 & $C_{3}H_{6} + * \leftrightarrow{}  C_{3}H_{6}*$ & -0.1 & 0.2 \\
& 4 & $C_{3}H_{8}* + O* \leftrightarrow{}  C_{3}H_{6}* + H_{2}O + *$ & -0.35 & 1.54\\
& 5 & $C_{3}H_{8}* + 2O* \leftrightarrow{}  C_{3}H_{4}* + 2H_{2}O + 2*$ & -3.98 & 1.65\\
& 6 & $C_{3}H_{6}* + O* \leftrightarrow{}  C_{3}H_{4}* + H_{2}O + *$ & -3.62 & 1.37 \\
\rot{\rlap{~Mechanism \#{} 1}}
& 7 & $C_{3}H_{4}* + 8O* \leftrightarrow{}  3CO_{2} + 2H_{2}O + 9*$ & -8 & 0.1 \\
\hline
& 1 & $C_{3}H_{8} + * \leftrightarrow{}  C_{3}H_{8}*$ & -0.2 & 0.3 \\
& 2 & $O_{2} + 2\wedge{} \leftrightarrow{}  2O\wedge{}$ & -0.7 & 1.25\\
& 3 & $C_{3}H_{6} + * \leftrightarrow{}  C_{3}H_{6}*$ & -0.1 & 0.2\\
& 4 & $C_{3}H_{8}* + O\wedge{} \leftrightarrow{}  C_{3}H_{6}* + H_{2}O + \wedge{}$ & -0.35 & 1.54\\
& 5 & $C_{3}H_{8}* + 2O\wedge{} \leftrightarrow{}  C_{3}H_{4}* + 2H_{2}O + 2\wedge{}$ & -3.98 & 1.65\\
& 6 & $C_{3}H_{6}* + O\wedge{} \leftrightarrow{}  C_{3}H_{4}* + H_{2}O + \wedge{}$ & -3.62 & 1.37\\
\rot{\rlap{~Mechanism \#{} 2}}
& 7 & $C_{3}H_{4}* + 8O\wedge{} \leftrightarrow{}  3CO_{2} + 2H_{2}O + * + 9\wedge{}$ & -8 & 0.1\\
\hline
& 1 & $C_{3}H_{8} + * \leftrightarrow{}  C_{3}H_{8}*$ & -0.2 & 0.3 \\
& 2 & $C_{3}H_{8} + \wedge{} \leftrightarrow{}  C_{3}H_{8}\wedge{}$ & -0.7 & 1.25\\
& 3 & $O_{2} + 2* \leftrightarrow{}  2O*$ & -0.1 & 0.2\\
& 4 & $C_{3}H_{6} + * \leftrightarrow{}  C_{3}H_{6}*$ & -0.35 & 1.54\\
& 5 & $C_{3}H_{8}\wedge{} + O* \leftrightarrow{}  C_{3}H_{6}* + H_{2}O + \wedge{}$ & -3.98 & 1.65\\
& 6 & $C_{3}H_{8}* + 2O* \leftrightarrow{}  C_{3}H_{4}* + 2H_{2}O + 2*$ & -3.62 & 1.37\\
\rot{\rlap{~Mechanism \#{} 3}}
& 7 & $C_{3}H_{6}* + O* \leftrightarrow{}  C_{3}H_{4}* + H_{2}O + *$ & -8 & 0.1\\
& 8 & $C_{3}H_{4}* + 8O* \leftrightarrow{}  3CO_{2} + 2H_{2}O + 9*$ & 0.36 & 1.45\\
\hline

\end{tabular}
\end{center}
\end{table}

\subsection{Workflows for MBDoE of TAP experiments}

\begin{figure}[H]
  \includegraphics[width=18cm]{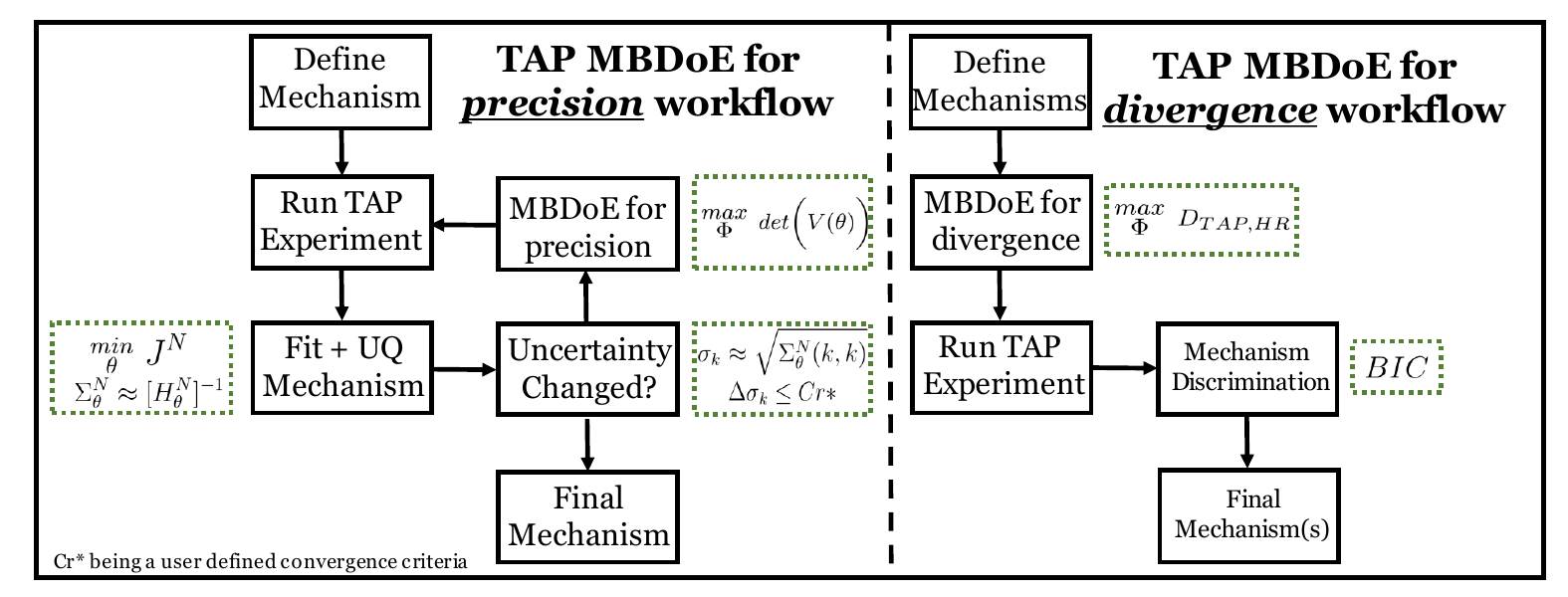}
  \caption{The proposed general workflow for selecting additional TAP experiments, where the predicted optimal experiment builds from the model-based design of experiments. }
\label{mbdoe-workflow}
\end{figure}

In this study, we introduce two general workflows for reducing parametric uncertainty (precision) and structural uncertainty (divergence) in kinetic models using TAP experiments. The workflows are outlined in Figure \ref{mbdoe-workflow}. We propose the user begin with a potential reaction mechanism. In the case of precision refinement, the user would then run an arbitrary TAP experiment. Following the experiment, the initial round of optimization and uncertainty quantification (UQ) should be performed. Next, the optimality criteria should be used to determine the impact the experimental conditions have on parameter identifiability. If high variation is observed, the experiment with the highest value for the criteria should be selected, performed, and re-optimized. This continues until the user has sufficiently reduced the uncertainty or the predicted optimal criteria begins to experience limited reduction. In the case of divergence, it is assumed that the different possible mechanisms and parameters can be defined a priori. From there, the MBDoE for divergence will identify the experiment that maximizes the divergence between the mechanisms, and fitting the different mechansims to the results of the experiment enables mechanism discrimination by comparing an information criterion. Details and variations of these workflows are described in detail in the subsequent sections.

\subsection{MBDoE for precision in TAP}

TAP experiments allow flexible specification of initial conditions, including pulse intensity, pump/probe delay, surface coverage, and reactor temperature. The feed concentrations (i.e. intensities) can be readily changed in any experiment, providing the experimentalist with strong control over surface coverage variations. The tuning of these parameters allows for the targeted expression of elementary processes in a reactor system, which can be beneficial to fully understand the mechanism and intrinsic kinetic parameters. The understanding of a mechanism is not static and can evolve with the inclusion of new experimental data. Identifying which experiments will yield the most insight into catalytic systems can be complex and relying on chemical intuition can be inefficient. MBDoE offers a quantitative strategy for the selection of additional experiments that maximize information content.

The first step in designing experiments is to arrive at an initial guess of the parameters, optimize them, and quantify the associated uncertainties (Fig. \ref{mbdoe-workflow}). Using this initial understanding of the model and parametric uncertainty, it is possible to predict the conditions that will reduce the uncertainty the most in future experiments. This is achieved through the use of output sensitivities with respect to parameters of interest over time, or the dynamic sensitivity matrix. The dynamic sensitivity matrix is constructed as: 

 \begin{equation}
  \mathbf{Q}_{i} = \begin{bmatrix}
     \left.\frac{\partial{}\hat{f}_{i}}{\partial{}\theta_1}\right\rvert_{\textrm{t}} & ... & \left.\frac{\partial{}\hat{f}_{i}}{\partial{}\theta{}_{m}}\right\rvert_{\textrm{t}}\\
   \vdots & \ddots & \vdots\\
 \left.\frac{\partial{}\hat{f}_{i}}{\partial{}\theta_1}\right\rvert_{\tau{}} & ... & \left.\frac{\partial{}\hat{f}_{i}}{\partial{}\theta{}_{m}}\right\rvert_{\tau{}}
 \end{bmatrix}=\begin{bmatrix}\nabla_{\boldsymbol{\theta}}{f}^T_i(t)\\\vdots\\\nabla_{\boldsymbol{\theta}}{f}_i^T(\tau)\end{bmatrix}\label{dynamicSensitivity}
 \end{equation}   

with $\hat{f}_{i}$, $\theta{}$, $m$, and $\tau$ representing the simulated outlet flux, the fitted parameter of interest, the final parameter being considered in the set, and the total number of time steps in the system, respectively. A matrix is constructed for each outlet gas $i$ and has a number of columns equal to the number of parameters and a number of rows equal to the number of time steps ($\tau{}$). These dynamic matrices are then consolidated into Fisher information matrices \cite{fisher1937design}. 
The Fisher information matrix quantifies how informative a new experiment will be with respect to a proposed kinetic model and the associated parameters. The Fisher information matrix is defined in terms of the dynamic sensitivity matrix as:
  \begin{equation}\label{fisherInformationMatrix}
    \mathbf{V}(\boldsymbol{\theta}) = \bigg[ \sum^{G}_{i}\sum^{G}_{j}\sigma{}_{i,j}^{2}\mathbf{Q}_{i}^{T}\mathbf{Q}_{j} + \left(\operatorname{\Sigma}^{N}_{\boldsymbol{\theta}}\right)^{-1} \bigg]^{-1} 
  \end{equation}

where $\Sigma{}^{N}_{\theta}$ is the covariance matrix of the model (see Equations \ref{objectiv_function_std} through \ref{hess3}), $\sigma{}_{i,j}$ is the outlet flux noise, and $i$ is a species in the set of gasses $G$. The Fisher information matrix is typically distilled into a single quantitative criterion. Interpreting and comparing scalar values is simpler than comparing matrices. Three criteria for optimality are frequently used in the MBDoE for precision. One is the A-optimality, which is a measure of the trace of the information matrix (or a sum of the total variances):

\begin{equation}
    A_{opt} = \operatorname{tr}( \mathbf{V}(\boldsymbol{\theta}{}) )
\end{equation}

Another is the E-optimality, which aims to minimize the largest eigenvalue of the covariance matrix:

\begin{equation}
    E_{opt} = \operatorname{max}[\operatorname{eig}( \mathbf{V}(\boldsymbol{\theta}) ]
\end{equation}

Finally, the D-optimality is the determinant of the Fisher information matrix (or a product of the total variances):

\begin{equation}
    D_{opt} = \operatorname{det}( \mathbf{V}(\boldsymbol{\theta}) )
\end{equation}

When trying to predict an optimal experiment, it is desired to find the experiment that will result in the lowest A, D, or E-value. These minimum values are meant to correlate with the maximum amount of information in the system, where each criterion corresponds to a different definition of information.

We first evaluate MBDoE as a strategy to systematically reduce the uncertainty on fitted parameters. We utilize the parameters for Mechanism 1, as defined in Table \ref{opdh_model_1} for all data generation in this section. We focus the analysis on a sub-set of 7 parameters ($\Delta G_0$, $\Delta G_1$, $\Delta G_2$, $G^{\ddagger}_1$, $G^{\ddagger}_3$, $G^{\ddagger}_4$, $G^{\ddagger}_6$), selected based on an initial sensitivity analysis which identifies these as the parameters that have the most influence on the model (provided in the SI). This is consistent with the similar energy scale for these parameters ($\Delta G \sim (-1,0) $ eV, $G^{\ddagger} \sim (1,2)$ eV), since very low barriers lead to effectively equilibrated reactions and very negative reaction free energies lead to effectively irreversible steps, making it difficult to deduce these parameters from experimental data. For the inverse problems in this work, we fix the excluded parameters to their true value, and use initial guesses of $-0.3$~eV for all adsorption/reaction energies and $1.5$~eV for activation energies. In practice, determining these initial guesses would require some prior knowledge, global optimization techniques, and sensitivity analyses. However, here we are primarily concerned with the ability of MBDoE to refine the precision of parameter estimates and restrict our analysis to how additional experimental data improves accuracy and reduces uncertainty on local optimization of kinetic parameters.

\subsubsection{MBDoE for divergence in TAP}

\begin{figure}[H]
  \includegraphics[width=11cm]{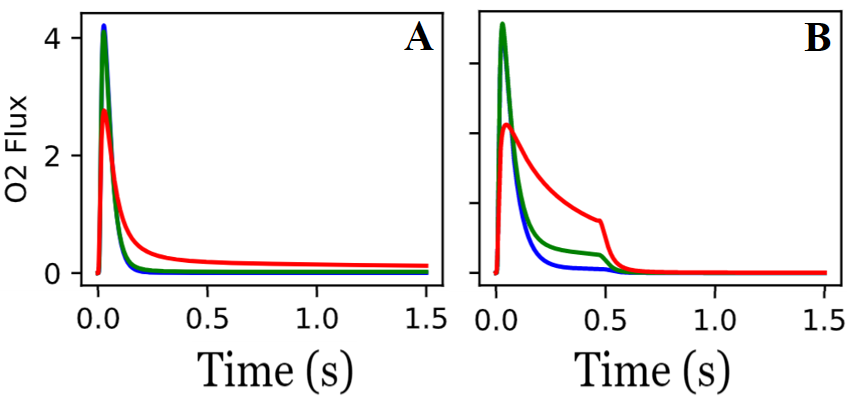}
  \caption{An example of mechanism divergence observed between three potential mechanism (green, blue, and red) simulations at two different TAP experimental conditions (subplots A and B). The two experiments do not equally differentiate the mechanisms, with experiment B showing a stronger divergence between the green and blue mechanisms in the tail of the outlet flux.}
\label{divergence_example}
\end{figure}

The  goal of MBDoE for model divergence is to identify a crucial experiment that can be used to differentiate between different possible models that describe the data, as illustrated in Figure \ref{divergence_example}. To achieve this, each known model is used to simulate the system over a wide range of initial condition space. At each combination of initial conditions selected, the simulated outlet flux will be available for all gas-phase species for each of the mechanisms being explored. Of the initial conditions explored, the one that maximizes the divergence between the results of the simulations from different models is selected. When there is no consideration of the uncertainty, the divergence criteria can be written in a form derived by Hunter and Reiner and adjusted specifically for TAP as:

\begin{equation}\label{divergenceCriteria}
\begin{split}
   D_{TAP,HR} &= \sum^{M-1}_{j=1}\sum^{M}_{k=j+1}\sum^{\tau{}}_{t=1}{\boldsymbol{\varepsilon}_{t,k,j}^T\boldsymbol{\Omega}_t\boldsymbol{\varepsilon}_{t,k,j}}\\
   \text{for}&\:\boldsymbol{\varepsilon}_{t,k,j}=\hat{\mathbf{f}}_{t,j} - \hat{\mathbf{f}}_{t,k}\text{, and}\:\boldsymbol{\Omega}_t=\operatorname{diag}(\{\sigma_{i}(t)^{-2}\})
 \end{split}     
\end{equation}

where $M$ is our set of models, $G$ is the set of gasses in our system, $\tau{}$ is the time steps where flux data points were collected, and $\hat{\mathbf{f}}=\{\hat{f}_i\}$ is the simulated flux \cite{hunter1965designs}. 

Discriminating between mechanisms based on qualitative deviations can be challenging and lead to varying results based on the user's intuition \cite{olofsson2019gpdoemd}. For this reason, significant efforts have been made to establish quantitative criteria for selecting between competing mechanisms. We evaluated the quality of a model using the Bayesian information criteria (BIC), which penalizes models with more parameters \cite{neath2012bayesian}. BIC is a commonly used method for selecting between competing mechanisms and is defined as:

\begin{equation}
    BIC = kln(n) - 2ln(J^{n})
\end{equation}

with $n$ representing the sample size (number of data points) of the system and $k$ representing the number of parameters. BIC parameter penalization increases with the size of the data set and the values are evaluated with kinetic parameters in the following section. We note that the BIC can be calculated with any set of kinetic parameters, and in the subsequent section and discussion the BIC with non-optimized and optimized kinetic parameters are compared. 

\section{Results}
\subsection{MBDoE for precision}

To initiate MBDoE, we first need an  experiment to establish parameter estimates.
A simple choice for an initial experiment is equimolar pulses of propane and oxygen. Co-pulsing these species simultaneously is selected since it is a natural boundary in the delay space. A reactor temperature of $700$ K was selected as a typical temperature of an OPDH reaction. The experiment was simulated at these conditions and the kinetic parameters of the model were optimized to the data set using standard TAPsolver settings and initial guesses as described in the methods section. The fitted model is presented in Figure \ref{opdh_initial_fit_precision} and the optimized parameters and their $95$\% confidence intervals are shown in Table \ref{just_parameters}. The results reveal strong agreement between most of the fitted parameters and the ground truth values. Most of the fitted parameters are within $0.02$ eV of the ground truth, with the exception of the free energy of oxygen adsorption ($\Delta{}G_{1}$) which has a more significant error of $0.17$ eV. The uncertainty in $\Delta{}G_{1}$ is likely the result of a low sensitivity of the parameter to the experimental conditions.

The confidence intervals of many parameters are on the order of $0.01$ - $0.1$ eV. Given that rates depend exponentially on these free energy values, it is of interest to systematically reduce the confidence intervals to yield more precise estimates.

\begin{table}[H]
\begin{center}
\small\addtolength{\tabcolsep}{-5pt}
\caption{The actual values, initial guesses, and values following optimization with simulated experimental data sets 1, 2, and 3 (defined in S\ref{just_exper}). The confidence intervals (95\%) of the parameters following optimization are also provided. The values of the alternative approach (labeled Exp. Alt.) is also presented in the right most column.}\label{just_parameters}
\begin{tabular}{|c@{\hskip 0.08in}|c@{\hskip 0.11in}|c@{\hskip 0.1in}c@{\hskip 0.17in}|c@{\hskip 0.14in}c@{\hskip 0.17in}|c@{\hskip 0.14in}c@{\hskip 0.17in}|c@{\hskip 0.14in}c@{\hskip 0.17in}|c@{\hskip 0.13in}c@{\hskip 0.13in}|}
\hline
& Actual & \multicolumn{2}{c|}{Initial} & \multicolumn{2}{c|}{Exp. 1}  & \multicolumn{2}{c|}{Exp. 2}  & \multicolumn{2}{c|}{Exp. 3} & \multicolumn{2}{c|}{Exp. Alt.} \\
\hline
 & Value & Value & C.I. & Value & C.I. & Value & C.I. & Value & C.I. & Value & C.I. \\
\hline
$\Delta{}G_{0}$ & -0.20 & -0.30 & \textbf{--} & -0.20 & 2.73e-3 & -0.20 & 8.17e-4 & -0.20 & 5.77e-4 & -0.20 & 6.14e-4\\
$\Delta{}G_{1}$ & -0.70 & -0.30 & \textbf{--} & -0.53 & 9.51e-2 & -0.70 & 2.25e-2 & -0.71 & 2.19e-2 & -0.70 & 2.67e-3\\
$G^{\ddagger}_{1}$ & 1.25 & 1 & \textbf{--} & 1.24 & 2.57e-2 & 1.24 & 5.98e-4 & 1.25 & 4.26e-4 & 1.25 & 5.04e-4\\
$\Delta{}G_{2}$ & -0.10 & -0.30 & \textbf{--} & -0.10 & 4.40e-2 & -0.12 & 9.27e-3 & -0.10 & 1.03e-2 & -0.12 & 6.73e-3\\
$G^{\ddagger}_{3}$ & 1.54 & 1.5 & \textbf{--} & 1.54 & 2.41e-3 & 1.54 & 8.49e-4 & 1.54 & 6.09e-4 & 1.54 & 6.67e-4\\
$G^{\ddagger}_{4}$ & 1.64 & 1.5 & \textbf{--} & 1.63 & 3.88e-2 & 1.63 & 5.94e-3 & 1.65 & 6.55e-3 & 1.63 & 5.25e-3\\
$G^{\ddagger}_{6}$ & 1.37 & 1.5 & \textbf{--} & 1.37 & 4.51e-2 & 1.39 & 9.47e-3 & 1.37 & 1.05e-2 & 1.39 & 6.79e-3 \\
\hline
\end{tabular}
\end{center}
\end{table}

The first round of MBDoE can be performed once initial parameter estimates and uncertainties are established. We perform a grid search over the initial conditions considered and calculate the Fisher information matrices (Eq. \ref{fisherInformationMatrix}). Next, an optimality criterion must be selected. Based on prior literature, we select D-optimality as the criterion to use, so the optimal experiment is the one with the smallest determinant of the inverse Fisher information matrix \cite{waldron2019autonomous,waldron2020model}. We also explored other criteria and confirmed that D-optimality is the most effective in this case (see SI). Using the D-optimality criterion, we find that an experiment with 2 nmol of propane and oxygen pulses, a propane delay of 0.6 seconds, and a temperature of 650 K was best in the grid search (a table of the designed experiments are shown in the SI). We re-fit the parameters with the data from this new experiment and the data from the first experiment included in the loss function. The new values of the parameters and their confidence are presented in Table \ref{just_parameters}. The estimates of most parameters remain approximately constant, with the exception of $\Delta{}G_{1}$, which exhibits strong agreement with the ground truth and significantly lower error bars after the addition of the new data. The confidence intervals of other parameters also decrease, indicating that the additional experiment indeed improves the accuracy and precision of the estimated parameters.

Following the MBDoE for precision workflow outlined in Figure \ref{mbdoe-workflow}, we continue with another iteration of the MBDoE to see if the uncertainty can be further reduced. We follow the same grid search approach previously used and find that a similar experiment is identified. However, the delay is selected to be 0.15 seconds instead of 0.6 seconds. We add the data from this additional experiment to the loss function and provide the parameter values and confidence intervals after optimization in Table \ref{just_parameters}. Both the parameter estimates and confidence intervals are relatively similar after including this additional data set. Some confidence intervals are slightly reduced, while others see a slight increase, which could be a result of variations in the confidence intervals of other parameters. The approximately static parameter estimates and confidence intervals suggest that the MBDoE workflow has converged and that further experiments are unlikely to yield additional improvements in precision.

\begin{figure}[H]
    \centering 
  \includegraphics[width=16cm]{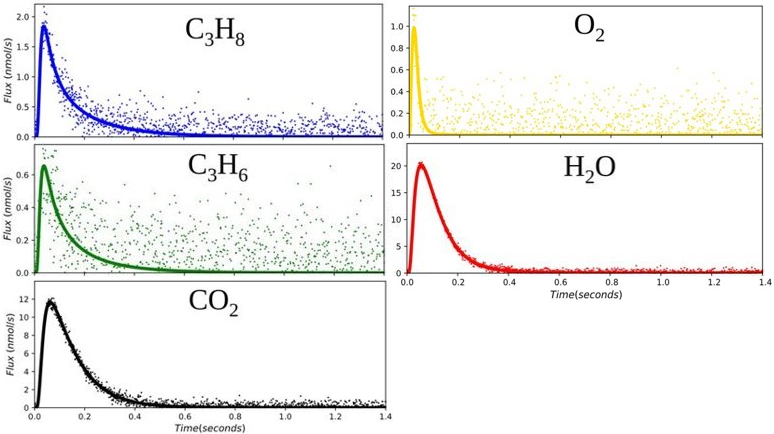}
  \caption{Initial OPDH mechanism fit (model results are represented by each line, while data is represented by the points) to the synthetic experimental data using a reactor temperature of 700 K, a 1 nmol pulse of $C_{3}H_{8}$ and $O_{2}$, and no delay between either gas pulse.}
  \label{opdh_initial_fit_precision}
\end{figure}

\subsection{MBDoE for model discrimination}

The use of MBDoE for model discrimination follows a different structure and set of assumptions from MBDoE to improve parameter precision. 
We use the mechanisms and associated parameters found in Table \ref{opdh_model_1}, but in this case the goal is to investigate whether TAP experiments can be used to distinguish these subtly different mechanisms. To create a more realistic scenario, we add a small amount of Gaussian noise ($\sigma = 0.05$ eV) to the parameters for each simulated experiment to mimic the effect of small variations between different experiments (e.g. different active site concentrations).
In this scenario, we assume that three possible mechanisms and associated rate constants have been identified, and the goal is to determine the  conditions of the TAP experiment that lead to the greatest divergence between the results of experiments for different mechanisms. In principle, this leads to three different ``ground truths'', corresponding to scenarios where each of the possible mechanisms is the true one. Here, we focus on the case where mechanism 2 is the ``ground truth'' since it yielded results where model discrimination was non-trivial (see SI).

\begin{figure}[H]
    \centering 
  \includegraphics[width=10cm]{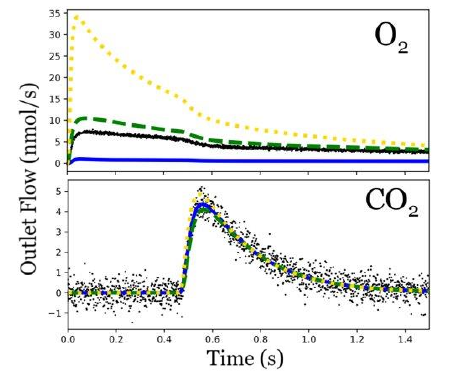}
  \caption{A visual of the differences observed between mechanism 1 (solid blue), mechanism 2 (dashed green), and mechanism 3 (dotted yellow) for the outlet flux of oxygen and carbon dioxide. Propane, propene, and water had agreements similar to those found in the carbon dioxide subplot above, while oxygen was the only graph with significant degrees of divergence. Mechanism 2 was found to agree most with the experimental data (black dots) and was quantitatively confirmed with the BIC values.}
  \label{det_flux}
\end{figure}

The same grid search approach over different experimental conditions is applied, as previously introduced for MBDoE for precision. However, in this case there is no parameter estimation or uncertainty quantification -- the goal is to maximize the difference between each of the outlet fluxes of each mechanism. When running this analysis, we found the set of conditions that led to the highest divergence to be 2 nmol of propane and oxygen, a propane delay of 0.45 seconds, and a temperature of 700 K. All fluxes, with the exception of oxygen, were found to agree with the experimental data reasonably well. For this reason, only oxygen and carbon dioxide are presented in Figure \ref{det_flux}. In the oxygen flux subplot, mechanisms 1 and 3 (solid blue and yellow, dotted line, respectively) are clear outliers and can be discarded. Mechanism 2 (dashed, green line) falls within a similar window of outlet flux values. Mechanism 2 agrees more with the simulated experimental data and has a lower BIC, indicating it is the most likely mechanism.  These results indicate that TAP experiments are capable of discriminating between subtle differences in reaction mechanisms, even when the underlying kinetic parameters are very similar, and that MBDoE is an effective route to identifying which TAP experimental conditions are expected to provide the most discrimination between different candidate mechanisms.

\section{Discussion}

\subsection{Efficacy of MBDoE for Precision}

\begin{figure}[H]
    \centering 
  \includegraphics[width=14cm]{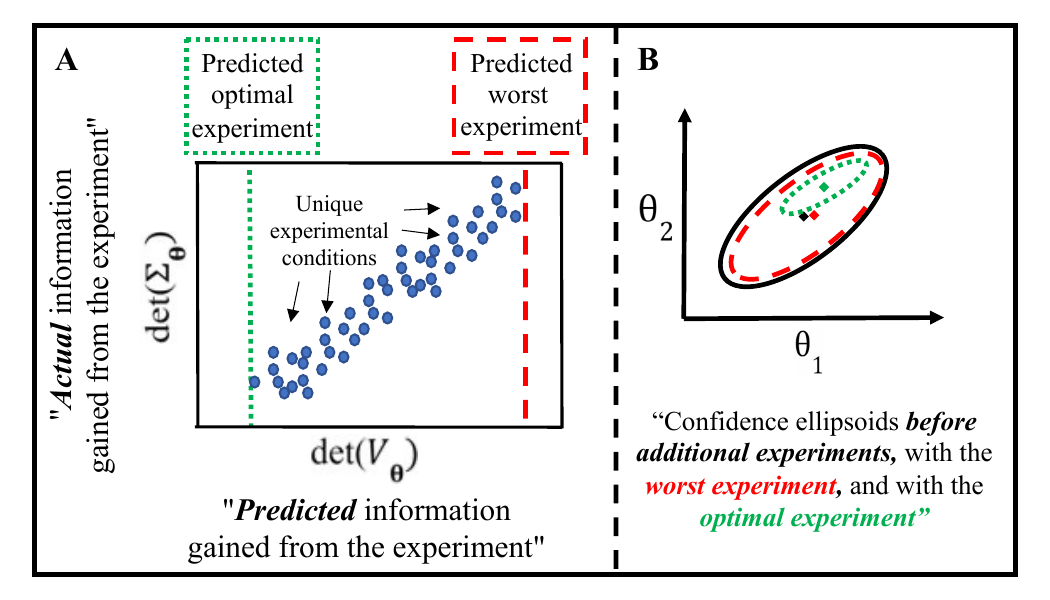}
  \caption{The correlation between predicted information gain (determinant of the inverted Fisher information matrix in D-optimal design) vs. the actual information gain (the determinant of the covariance matrix for re-fitted model in D-optimal design) for a given parameter reveals the ability of MBDoE to identify the optimal experiment, and the impact of using the optimal experiment.}
  \label{precision_concept}
\end{figure}

The results section shows that the experiments selected by MBDoE reduce the uncertainty on fitted parameters. 
However, the efficacy of MBDoE over random  experimentation is not clear. To provide a more rigorous evaluation of the performance of MBDoE, we explored the correlation between predicted and actual information gained, where the predicted information is defined by the determinant of the inverse Fisher information matrix and the actual information is defined by the determinant of the covariance matrix after re-fitting the model \cite{franceschini2008model}. This is conceptually presented in Figure \ref{precision_concept}. Importantly, the Fisher information matrix is available without generating additional synthetic experimental data, while the covariance matrix requires experiments to be run/simulated. Thus, the comparison between predicted and actual information is only practical where (synthetic) experimental data can be easily generated, and is used here to analyze the performance in a simulated scenario where this is possible.

The correlation in these graphs indicates the accuracy of the predicted information, while the difference between the minimum and maximum actual information indicates the influence of the experimental conditions on a given parameter, so these graphs provide a convenient visual approach to evaluating the efficacy of MBDoE for precision. We used this approach to evaluate the choice of optimality criterion, comparing predicted vs. actual information for D-, E-, and A-optimal criteria. The results, shown in the SI, show that the correlation is strongest for D-optimal criterion. This justifies the use of the D-optimal criterion in this work, and we focus on D-optimal experiments in all subsequent analyses.

\newpage

\begin{figure}[H]
    \centering 
  \includegraphics[width=13.5cm]{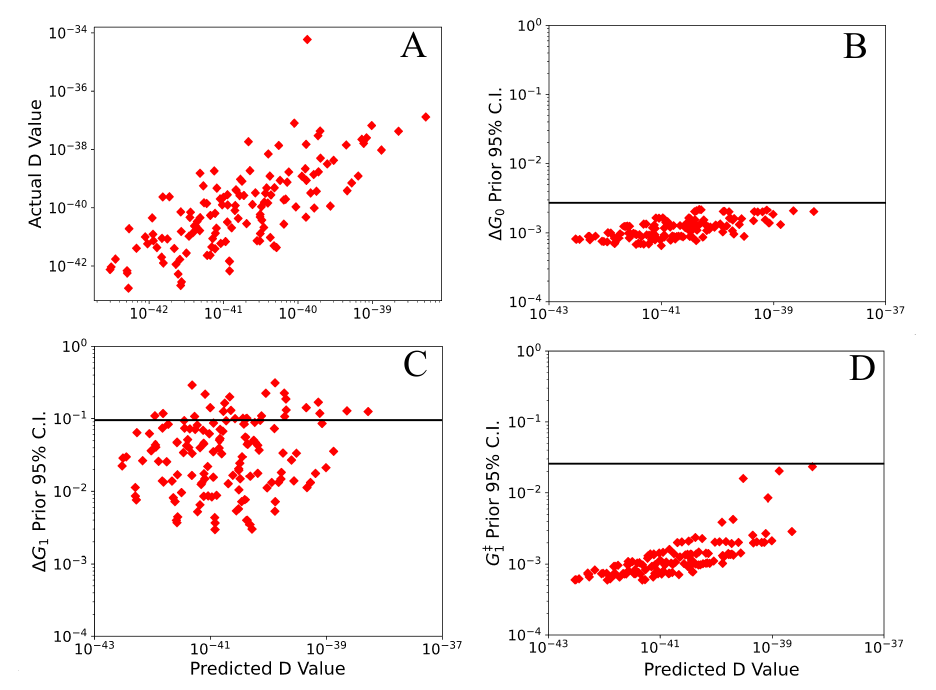}
  \caption{The predicted information (Predicted D-value) gain compared to the actual information gained (Actual D-value), as well as the confidence intervals for the parameters fitted in the system, for the second experiment. Each red point represents a different set of experimental conditions evaluated, while the black horizontal line represents the confidence interval found after the first experiment.}
  \label{optimal_std_1}
\end{figure}

The trends in the predicted information and the actual information after the first experiment are presented in Figure \ref{optimal_std_1}. Figure \ref{optimal_std_1}A shows the predicted D-optimal compared to the actual D-optimal (or determinant of the covariance matrix). There is a clear correlation between high and low D-optimal values, indicating that the determinant criterion should be a good predictor of parameter confidence interval reductions. Figure \ref{optimal_std_1}B shows the confidence intervals around the refitted values of $\Delta{}G_{0}$, or the free energy of adsorption of propane. There is a reasonable correlation between predicted and actual D-values, and the D-optimal experiment outperforms the majority of the competing experimental designs (approximately 96$\%$). Although the lowest confidence interval is not observed at the lowest predicted D-value, the overall reduction in uncertainty of $\Delta{}G_{0}$ is relatively small for all experiments. 

The activation of adsorption of oxygen ($G^{\ddagger}_{1}$)  was explored in Figure \ref{optimal_std_1}D and the uncertainty reduction found in $\Delta{}G_{0}$ is again observed. In this case, some experiments perform far worse than others, although no experiments lead to an increase in the uncertainty. The unfavorable experiments all fall at high predicted D-values and would be rejected. The uncertainty reduction in $G^{\ddagger}_{1}$ is also larger than most other parameters, reducing by more than an order of magnitude from the prior experiment. This again highlights the fact that the MBDoE approach performs differently depending on which parameter is being investigated.
 
Next, in Figure \ref{optimal_std_1}C, the free energy of oxygen adsorption ($\Delta{}G_{1}$) was explored. Unlike $\Delta{}G_{0}$ and all other parameters, there is no trend in this distribution, and the ``optimal experiment'' leads only to a moderate reduction in uncertainty. However, there are also examples of experiments where the confidence interval increases, indicating that while the D-optimal experiment may not be optimal, it is a significant improvement over randomly selecting experimental conditions.
On the other hand, some experiments lead to significantly more reduction in uncertainty than the predicted optimal experiment, indicating that it is possible to further reduce the uncertainty on this parameter, but that the MBDoE fails to identify the optimal experiment for this parameter. We hypothesize that this occurs due to the low sensitivity (high error bar) of $\Delta{}G_{1}$, which causes the MBDoE to favor improvement of parameters that are already well-determined; we revisit this issue later in the section.

The evaluation of other parameters --- propane adsorption ($\Delta{}G_{2}$), and activation energies for propane dehydrogenation, propane combustion, and propene combustion ($G_{3}^{\ddagger}$, $G_{4}^{\ddagger}$, and $G_{5}^{\ddagger}$) --- are also analyzed using the same technique. The results are shown in the SI, and are largely consistent with the findings for $\Delta{}G_{0}$. A reasonably strong correlation is observed between the predicted and actual D values, but the reduction in uncertainty is relatively small ($<1$ order of magnitude).

A similar analysis of predicted and actual D-values and confidence intervals for all parameters was also performed for the second iteration of MBDoE. The results, shown in the SI, show a much lower correlation between actual and predicted D-values, and significantly less reduction in uncertainty for the optimal experiment. Although some parameters show a slight decrease in uncertainty, most remain the same or even increase slightly. There are also more examples of experiments that lead to increased uncertainty for many parameters. These findings suggest that the precision of the parameters is saturated and that there is little value in additional experiments. This is consistent with the conclusion that the iterative MBDoE workflow shown in Figure \ref{mbdoe-workflow} has converged after two experiments.

\subsection{MBDoE for precision of specific parameters}

The results indicate that there is a lower limit to the precision that can be obtained by the standard MBDoE for precision workflow proposed in Figure \ref{mbdoe-workflow}. This may be problematic in the case where the uncertainties on one (or more) parameters of particular chemical interest are not sufficiently well determined after the workflow converges. 

The MBDoE for precision is typically used to reduce the uncertainty in all parameters simultaneously \cite{franceschini2008model}. Here, we propose a revised workflow that isolates a specific parameter and focus on the parameter with the highest degree of uncertainty, in this case $\Delta{}G_{1}$. We hypothesize that focusing on this parameter will lead to a more precise estimate of other parameters as well, since it will reduce the overall uncertainty in the system.

\begin{figure}[H]
    \centering 
  \includegraphics[width=13.5cm]{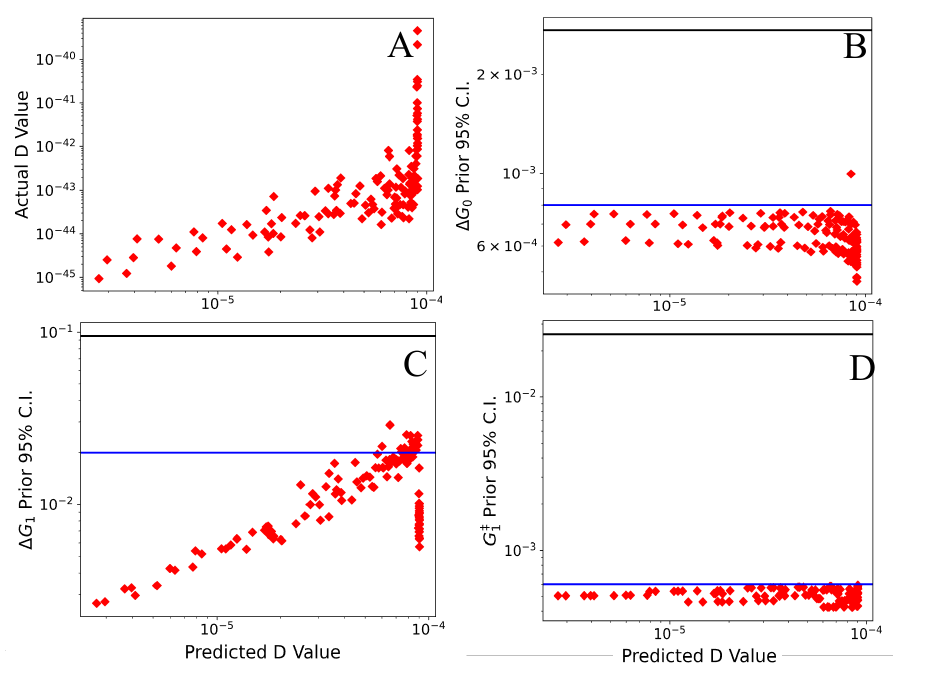}
  \caption{The predicted information (D-value) gain compared to the actual information gained (D-value), as well as the confidence intervals for the parameters fitted in the system, for the third experiment using only the parameter with the highest uncertainty in the design. Each red point represents a different set of experimental conditions evaluated, while the black horizontal line represents the confidence interval found after the first experiment, and the blue line represents the confidence interval found after the second experiment.}
  \label{optimal_std_alternative}
\end{figure}

To test this, we performed the MBDoE for precision, with only the $\Delta{}G_{1}$ value and uncertainty included in the grid search. We then analyzed the impact this selection criterion had on the actual information gained (optimizing all parameters at the experimental conditions) and confidence intervals for each parameter in the system. 
The results of this analysis are presented in Figure \ref{optimal_std_alternative}. The D-value for $\Delta{}G_{1}$ (Figure \ref{optimal_std_alternative}A)) shows a strong linear trend between the predicted optimal experiment and the true optimal experiment for most D-values, with a lack of correlation at high predicted D-values.

As expected, the uncertainty for $\Delta{}G_{1}$ (Figure \ref{optimal_std_alternative}C) is strongly correlated with the predicted D-value, with the optimal predicted experiment substantially reducing the uncertainty on $\Delta{}G_{1}$ by more than an order of magnitude.
The correlation between the D-value and the uncertainty for other parameters is much weaker, and experiments that were found to perform the worst for $\Delta{}G_{1}$ at times lead to the highest reduction in the uncertainty for other parameters. Specifically, $\Delta{}G_{0}$ (Figure \ref{optimal_std_alternative}B) and $G^{\ddagger}_{3}$ (shown in SI) had the lowest uncertainty for experiments that were predicted to be the worst for defining $\Delta{}G_{1}$, although other parameters (e.g. $\Delta G_{2}$ and $G^{\ddagger}_4$) had significantly increased uncertainty for the worst predicted experiment (shown in SI). These findings reveal that there can be anti-correlation between the optimality conditions for different parameters, providing insight into why the D-optimal design for all parameters simultaneously fails to systematically reduce the uncertainty on all parameters beyond a certain point.

On the other hand, the predicted optimal experiment for $\Delta{}G_{1}$ does lead to some reduction in the uncertainty for all parameters, and the resulting model has parameters that are more accurate and precise than the model where all parameters are included. These results indicate that focusing the DoE on a single poorly determined parameter is an effective strategy to reduce the uncertainty on that parameter, especially in the case that the full design of experiments does not significantly reduce uncertainty.

\subsection{Analysis of MBDoE for Divergence}
As with MBDoE for precision, we want to confirm that we are observing a trend between the predicted divergence and the ability to discriminate between the mechanisms. We performed a similar analysis as in the case of precision, simulating the experiments under all predicted conditions. In this case, we compare the predicted Hunter-Reiner divergence (Eq. \ref{divergenceCriteria}) with the BIC calculated for the various mechanisms compared to each experiment. The parameters are not re-optimized, and the spread in the BICs occurs due to the slight perturbations in kinetic parameters between experiments (as described in the methodology section). This analysis is presented in Figure \ref{disc_reoptimize}A. In all cases, the BIC of mechanism 3 is much higher than mechanisms 1 and 2, indicating that MBDoE is not necessary to discriminate between mechanism 3 and mechanisms 1 and 2. However, at low values of the mechanism divergence, there is a strong overlap between the BIC values for mechanisms 1 and 2, indicating that under these conditions it is not possible to determine which mechanism is consistent with the data. However, higher divergence values lead to significant differences in BIC values between mechanisms 1 and 2. This shows that TAP experiments are able to differentiate between different types of single and multi-site mechanisms, even when the underlying kinetic parameters are very similar. It also reveals that MBDoE for divergence is necessary to distinguish between subtly different multi-site mechanisms, while arbitrary experiments are sufficient for more distinct mechanisms such as single vs. multi-site mechanisms.

\subsection{Simultaneous determination of kinetic parameters and mechanism}

In the above analyses, it was assumed that the mechanism was known and the parameters were fitted and refined (precision), or that the parameters were known and the experiments were used to differentiate between different mechanisms (divergence). In practice, it is often the case that both the kinetic parameters and the reaction mechanism are unknown. Thus, in a realistic scenario, it would likely be necessary to combine these two workflows. For example, in the case of mechanism discrimination, each different mechanism might be re-fit to the experimental data. 

\begin{figure}[H]
    \centering 
  \includegraphics[width=15cm]{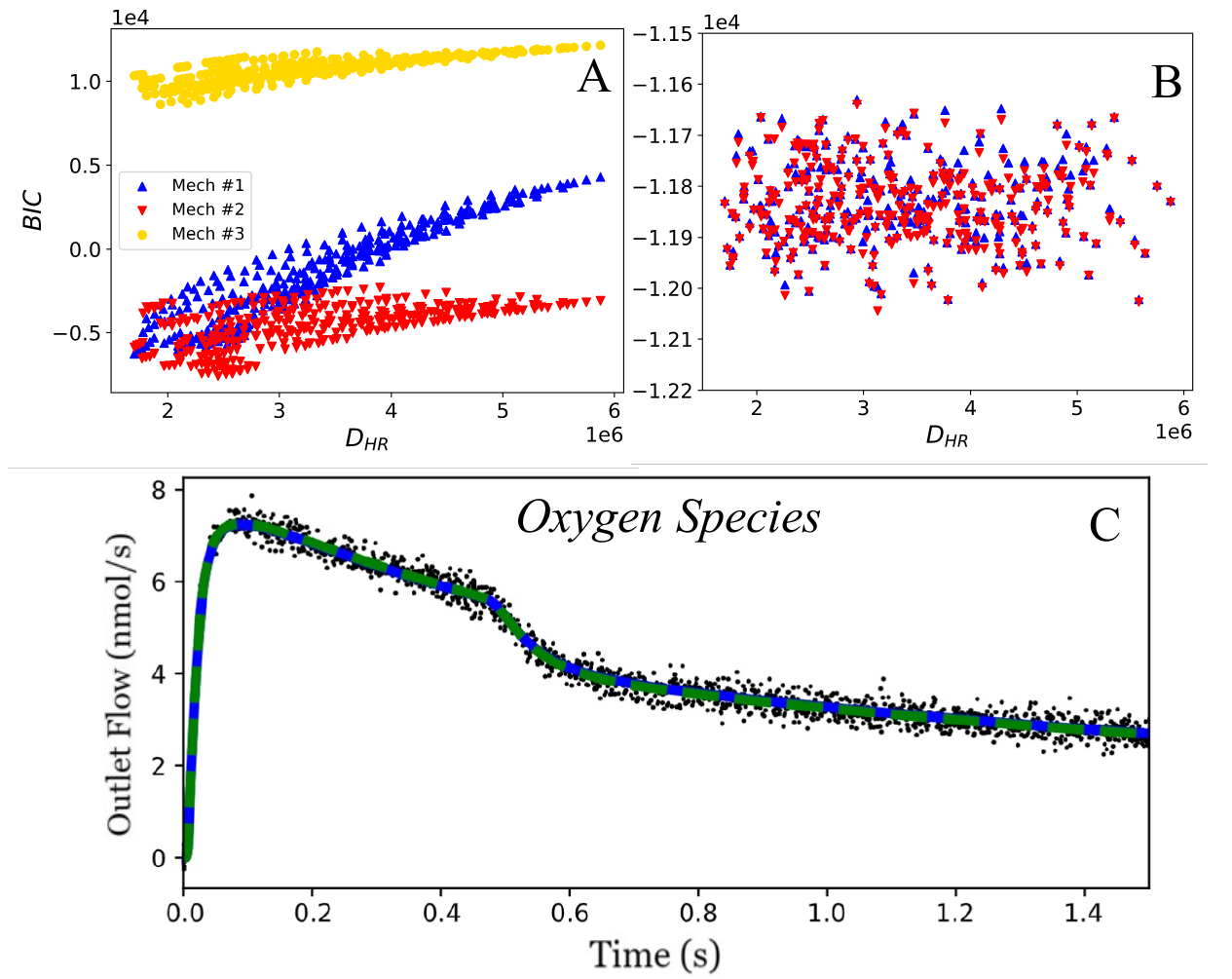}
  \caption{Subplot A shows the predicted divergence compared to the BIC value for each mechanism. At low predicted divergence values, it is more challenging to discriminate between mechanisms 2 and 3. At higher divergence values, mechanism 2 is highly favored (visualized in Figure \ref{det_flux}). Following optimization, the ability to adequately discriminate between mechanisms 1 and 2 is lost. Both mechanisms have similar BIC values (plot B) for each of the considered experimental conditions, with the outlet flow of oxygen (plot C) overlapping significantly.}
  \label{disc_reoptimize}
\end{figure}

To evaluate the combination of the proposed divergence and precision workflows, we re-optimize mechanisms 1 and 2 to the experimental data generated during the grid search for model divergence. This analysis is presented in Figure \ref{disc_reoptimize}B. Ideally, there would be some correlation between the predicted divergence and the difference between the BIC values for the two different mechanisms. It is clear that after reoptimization, the BIC is not correlated to the divergence, and discriminating between the two mechanisms is not possible under any of the experimental conditions explored. There are some experiments in which small differences are observable, but they are scattered randomly throughout the divergence and are too small to draw any strong conclusions. Prior optimal experiments based on the divergence criterion are also visually compared after re-optimization (Figure \ref{disc_reoptimize}C), and are essentially indistinguishable. These results suggest that the convolution of parametric and structural uncertainty presents a significant challenge since it is not possible to distinguish between mechanisms if the parameters are obtained by fitting to the kinetic data.   

The two components of this investigation, parameter fitting (precision) and mechanism discrimination (divergence) are often combined into a single workflow \cite{franceschini2008model}. A single mechanism is identified using discrimination approaches, or, more commonly, prior literature and intuition. Next, the parameters of this mechanism are fit, and can be refined using MBDoE.
 Our current results show that this approach becomes problematic when the flexibility in the mechanism increases, causing a scenario where complex mechanisms can describe a broad range of experimental data if parameter optimization is performed, even for transient kinetic datasets. This presents a challenge for the field, which could potentially be overcome with additional data in various forms. For example, multi-pulse (state-altering) experiments, or more data from experiments under diverse conditions, could be included to further constrain the parameters and mechanisms \cite{morgan2017forty}. Similarly, spectroscopic data can provide direct information on surface species, as well as on the structure of the catalytic material, potentially enabling better differentiation between different mechanisms and active site structures \cite{moncada2018developing}. Moreover, improvements in the optimization techniques may help alleviate the problem by providing more accurate and efficient estimates of parametric uncertainty \cite{gusmao2020kinetics,o2022derivate, Lejarza_2023}. Another approach is to explore alternate experimental design paradigms. Bayesian experimental design \cite{savara2020chekipeuq,walker2020chekipeuq} and robust information gain \cite{go2022robust} may enable the design of experiments that more effectively decouple parametric and structural uncertainty. Alternatively, modifications of the current workflow, such as reversing the order of mechanism divergence and precision and introducing additional experimental techniques, could also show promising results. For example, proposed mechanisms could be refined using TAP experiments and then validated (or discriminated between) using high pressure experiments \cite{wakefield2023fast,reece2021moving}. In addition, theoretical work to understand the limits of what parameters and mechanisms can be reliably identified from experimental data may help constrain candidate mechanisms to reduce flexibility and provide more efficient differentiation between candidate mechanisms \cite{roelant2010identifiability,redekop2014elucidating}.

\section{Conclusion}

Transient kinetic experiments provide investigators with dense data sets that can elucidate complex reaction mechanisms. The TAP reactor, a low-pressure transient kinetic method with millisecond time resolution, is particularly well-suited to deconvolute the intrinsic kinetics of catalytic materials \cite{morgan2017forty,yablonsky2003temporal}. The information gained from these experiments can be heavily dependent on the choice of initial conditions (e.g. pulse intensities, pump/probe delays, reactor temperature, etc.). Since no quantitative approach for identifying optimal TAP experiments has previously been introduced, we explore the use of MBDoE for precision and divergence in a synthetic oxidative propane dehydrogenation case study. 

The proposed workflow for selecting experiments and optimizing parameters is introduced and built around the Fisher information matrix, which combines current understanding of the parameters (covariance) and the predicted gain in information from a new experiment (the dynamic sensitivity matrices). We found that the MBDoE for precision is capable of identifying the most informative experimental conditions, resulting in an increase in confidence for most of the parameters. However, the reduction in uncertainty saturated after two experiments, and one parameter still had a significant confidence interval. A variation in the design criteria provided a route to further reduce the uncertainty of this parameter, as well as others, suggesting a clear path to refine the precision of fitted parameters through MBDoE.

Structural uncertainty, which is observed in catalytic studies in the form of active site configurations or inclusion of reaction steps, was also explored in TAP through MBDoE. The findings indicate that TAP is capable of differentiating between even subtle differences in reaction mechanisms, including different types of single and multi-site models. However, this differentiation relies on prior knowledge of the rate parameters that control the kinetics, and we find that it is not possible to differentiate between these subtly different mechanisms if parameter optimization is performed for each case. 

Additional complications may also arise when working with real experimental data. For example, certain molecules such as water may be difficult to observe or have a particularly high degree of uncertainty. Therefore, including the effects of missing or incomplete information in the design of experiments is expected to be an important area of future work. Generally, we expect that the MBDoE workflow will be an effective tool for guiding TAP experiments with quantitative feedback for both parameter refinement and model discrimination. Future implementations involving state altering experiments and spectroscopic data, as well as applications to non-Knudsen operating conditions could make this a more powerful tool, but improvements to efficiency and automation will be necessary.

\begin{acknowledgement}

Support for this work was provided by the U.S. Department of Energy (USDOE), Office of Energy Efficiency and Renewable Energy (EERE), Advanced Manufacturing Office Next Generation R\&D Projects under contract no. DE-AC07-05ID14517 and acknowledgement is made to the donors of the American Chemical Society Petroleum Research Fund for partial support of this research (61165-DNI5). This research made use of Idaho National Laboratory computing resources which are supported by the Office of Nuclear Energy of the U.S. Department of Energy and the Nuclear Science User Facilities under Contract No. DE-AC07-05ID14517.

\end{acknowledgement}

\newpage
\begin{suppinfo}

\setcounter{section}{0}
\setcounter{table}{0}
\setcounter{figure}{0}

\renewcommand\thefigure{\Alph{section}}
\renewcommand\thefigure{\Alph{figure}}
\renewcommand\thetable{\Alph{table}}

\section{Model parameter sensitivities for MBDoE for precision}
To narrow the parameters explored during the MBDoE for precision analysis, we looked at the sensitivity of all parameters to determine which are identifiable. Not all parameters will be experimentally observable due to their extreme values (i.e. if the energy is too high or too low, varying the energy in the model will not alter reaction rates). Using the initial parameter guesses and the seven-parameter fit, we show all the parameter sensitives in Table \ref{sensitivity_analysis}. The parameters excluded from the analysis have sensitives at and below 1e-3, whereas the included parameters had higher values. For this reason, we performed the analysis with these seven parameters.
\begin{table}[H]
\begin{center}
\small\addtolength{\tabcolsep}{-5pt} 
\caption{The sensitives of the parameters found with the initial parameter guesses and the final fit of the seven parameter optimization (the local minimum) . }\label{sensitivity_analysis}
\begin{tabular}{|c|c|c|}
\hline
Parameters & Initial Sensitivity & Local Minimum Sensitivity \\
\hline
$\Delta{}G_{0}$ & -4.92e5 & 7.64e-1 \\
$G^{\ddagger}_{0}$ & -1.80e-3 & 1.20e-5 \\
$\Delta{}G_{1}$ & 3.28e5 & 1.15e-2 \\
$G^{\ddagger}_{1}$ & -5.97e1 & -1.31e-1 \\
$\Delta{}G_{2}$ & -1.53e4 & 7.31e-2 \\
$G^{\ddagger}_{2}$ & 3.55e-6 & 2.42e-6 \\
$\Delta{}G_{3}$ & -3.08e1 & 1.07e-3 \\
$\Delta{}G_{4}$ & 4.45e-29 & -1.96e-33 \\
$\Delta{}G_{5}$ & 4.51e-24 & 2.70e-25 \\
$G^{\ddagger}_{3}$ & -1.61e5 & 1.04e0 \\
$G^{\ddagger}_{4}$ & -4.99e4 & 1.08e-1 \\
$G^{\ddagger}_{5}$ & -1.12e5 & 1.99e-1 \\

\hline
\end{tabular}
\end{center}
\end{table}

\section{Correlation between predicted D and actual D criteria}
We compared different methods for distilling the Fisher information matrix and calculated covariance matrices as the A (trace), D (determinant), and E (eigenvalue) criteria. Although each is used in the literature, we observed the strongest correlation for the D-optimal criteria and therefore use it for MBDoE.
\begin{figure}[H]
    \centering 
  \includegraphics[width=15cm]{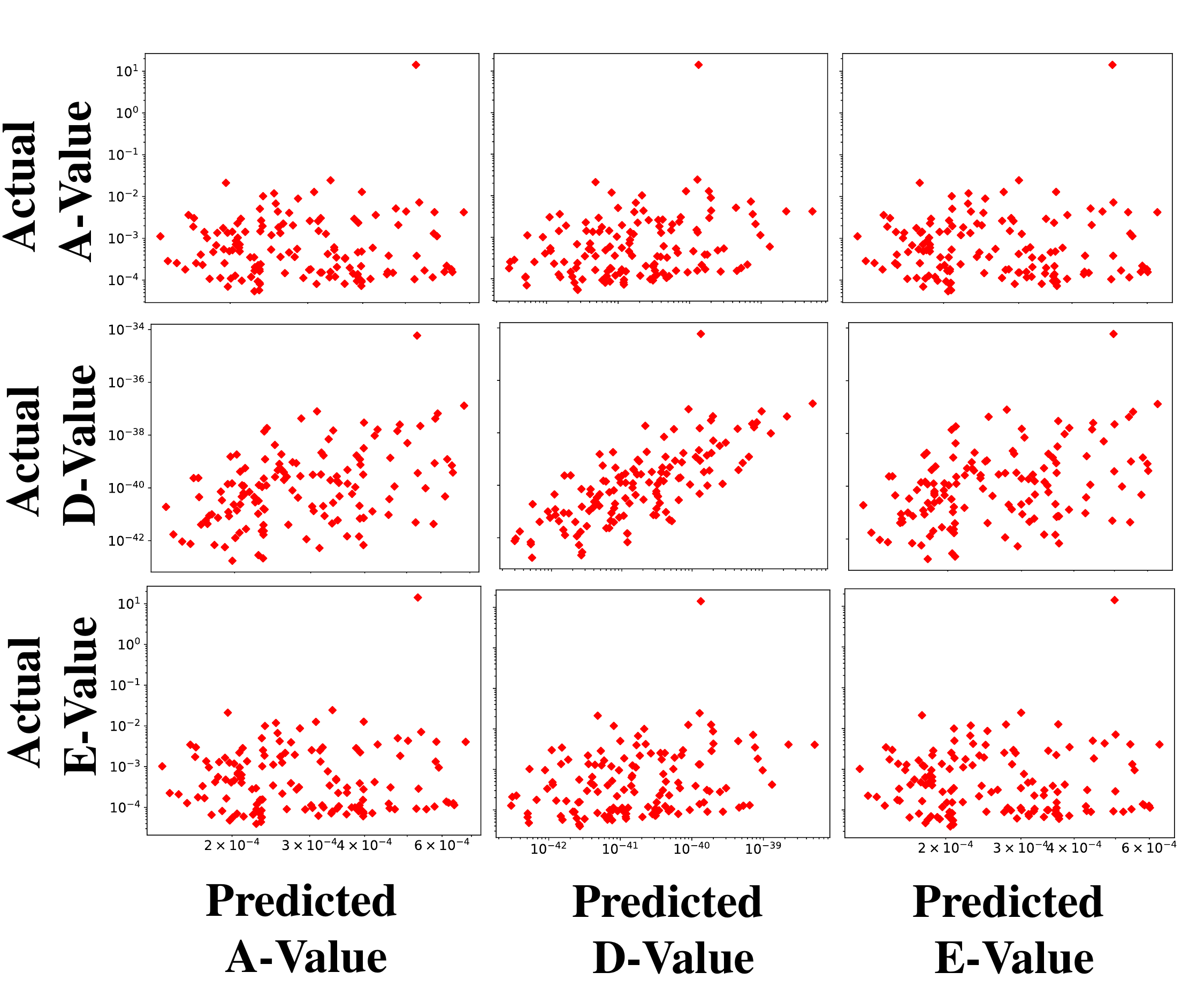}
  \caption{The predicted A, D, and E criteria for designing experiments were compared to their actual A, D, and E criteria for  each of the possible experiments. The D criteria, which stands for the determinant, is the only criteria that had a strong correlation and, for that reason, is the only criteria thoroughly explored in this paper.}
  \label{disc_disc}
\end{figure}
\newpage

\section{Initial conditions used during the precision refinement process}

The MBDoE approach for precision used an initial, arbitrary experiment ($1^{st}$ experiment in Table \ref{just_exper}), followed by three additional experiments. The $2^{nd}$ experiment is the first experiment predicted using MBDoE, while the $3^{rd}$ experiment is the second experiment predicted using MBDoE. The $Alt.$ experiment involves the adjusted approach for designing the experiments involving only the most uncertain parameter.

\begin{table}[H]
\begin{center}
\small\addtolength{\tabcolsep}{-5pt} 
\caption{The experimental conditions used to constrain the parameters. Experiment 1 was selected arbitrarily, while experiments 1 and 2 were selected through MBDoE for precision.}\label{just_exper}
\begin{tabular}{|c|c|c|c|c|}
\hline
& $C_{3}H_{8}$ Intensity (nmol) & $O_{2}$ Intensity (nmol) & $O_{2}$ Delay (s) &  Temperature (K)\\
\hline
$1^{st}$ Experiment & 1.0 & 1.0 & 0.00 & 700 \\
$2^{nd}$ Experiment & 2.0 & 2.0 & 0.60 & 650 \\
$3^{rd}$ Experiment & 2.0 & 2.0 & 0.15 & 650 \\
$Alt.$ Experiment & 2.0 & 2.0 & 0.60 & 700 \\

\hline
\end{tabular}
\end{center}
\end{table}
\newpage

\section{MBDoE experiment selection correlation}

We observed the performance of the MBDoE for several parameters in the oxidative propane dehydrogenation reaction, but focused only on three parameters in the primary text. For this reason, we provide the remaining plots in Figures \ref{additional_second} through \ref{additional_third} for the second experiment selection, third experiment selection, and the alternative third experiment selection. 

\begin{figure}[H]
    \centering 
  \includegraphics[width=15cm]{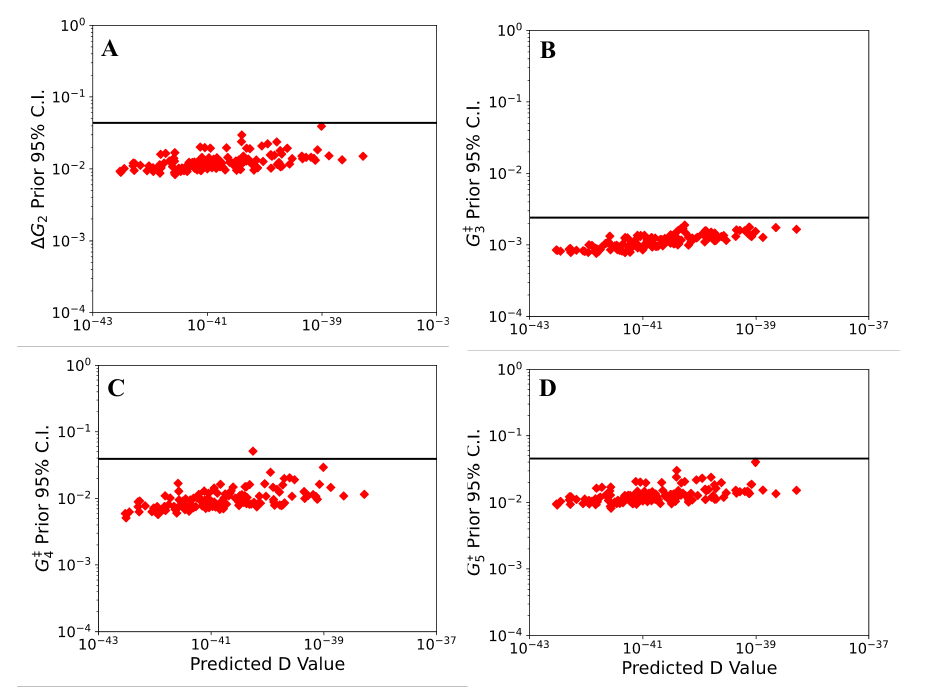}
  \caption{The predicted improvement to kinetic understanding using D-optimality for additional parameters fitted following the first experiment.}
  \label{additional_second}
\end{figure}
\newpage

\begin{figure}[H]
    \centering 
  \includegraphics[width=15cm]{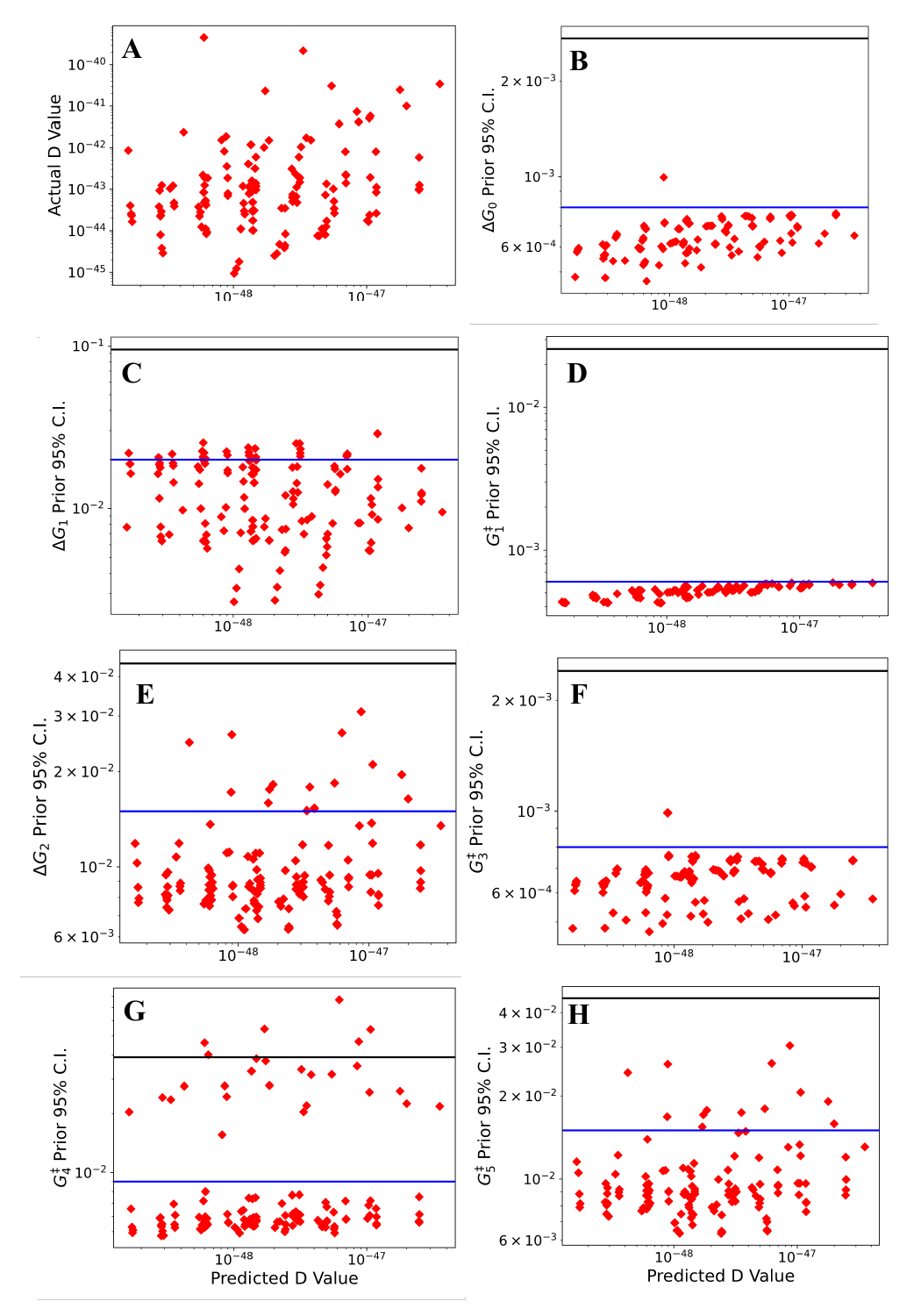}
  \caption{The predicted improvement to kinetic understanding using D-optimality for all parameters following the second experiment.}
  \label{all_third}
\end{figure}
\newpage

\begin{figure}[H]
    \centering 
  \includegraphics[width=15cm]{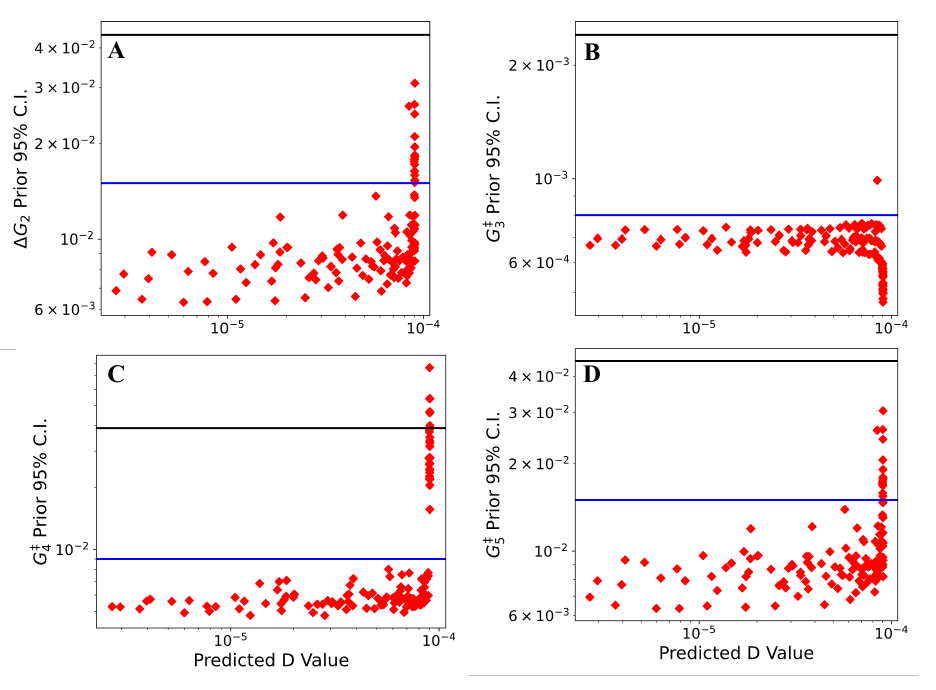}
  \caption{The predicted improvement to kinetic understanding using D-optimality for additional parameters fitted following the third experiment.}
  \label{additional_third}
\end{figure}
\newpage

\section{MBDoE experiment selection correlation}
Although we only considered mechanism two as the black box experiment in the primary text, we also explored the use of mechanism 1 and 3 as the black box experiment (shown in the top and bottom plots of Figure \ref{alt_divergence}). These experiments showed clear divergence between each other, so we focused on mechanism 2, which has non-trivial divergence for some experiments (e.g. there exist experiments for which it is not possible to clearly distinguish mechanism 1 and 2 based on BIC).
\begin{figure}[H]
    \centering 
  \includegraphics[width=8cm]{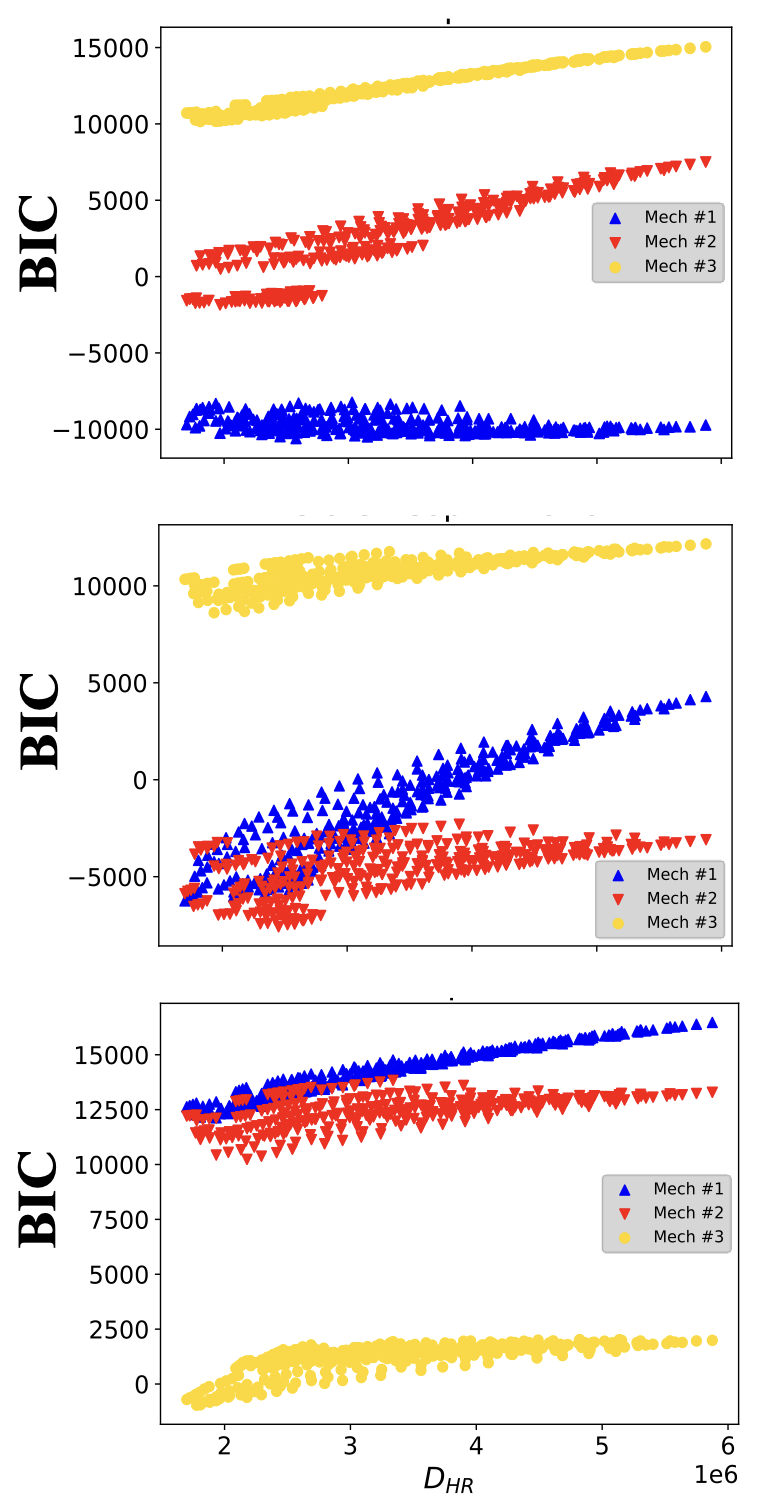}
  \caption{Divergence plots for the three oxidative propane mechanisms provided in the primary text. The top, middle, and bottom plots represent divergence when the experimental data is generated using mechanism 1, 2, and 3, respectively.}
  \label{alt_divergence}
\end{figure}
\newpage

\end{suppinfo}

\bibliography{achemso-demo}

\end{document}